\documentclass{elsart}
\usepackage{amsfonts}
\usepackage{amssymb}
\usepackage{amsmath}
\usepackage[dvips]{graphicx,psfrag,overpic,color}

\definecolor{dgreen}{rgb}{0,.6,0}
\usepackage[normalem]{ulem}

\begin{document}
\begin{frontmatter}

\title{A New Lorenz System Parameter Determination Method and
Applications}
\author[Spain]{A. Orue},
\author[Spain]{G. Alvarez\corauthref{corr}},
\author[Spain]{G. Pastor},
\author[Spain]{M. Romera},
\author[Spain]{F. Montoya} and
\author[China]{Shujun Li}

\corauth[corr]{Corresponding author: Email: gonzalo@iec.csic.es}

\address[Spain]{Instituto de F\'{\i}sica Aplicada, Consejo Superior de
Investigaciones Cient\'{\i}ficas, Serrano 144, 28006--Madrid, Spain}
\address[China]{Department of Electronic and Information Engineering,
The Hong Kong Polytechnic University, Hung Hom, Kowloon, Hong Kong
SAR, China}

\begin{abstract}
This paper describes how to determine the parameter values of the
chaotic Lorenz system from one of its variables waveform. The
geometrical properties of the system are used firstly to reduce the
parameter search space. Then, a synchronization-based approach, with
the help of the same geometrical properties as coincidence criteria,
is implemented to determine the parameter values with the wanted
accuracy. The method is not affected by a moderate amount of noise
in the waveform. As way of example of its effectiveness, the method
is applied to figure out directly from the ciphertext the secret
keys of two-channel chaotic cryptosystems using the variable $z$ as
a synchronization signal, based on the ultimate state projective
chaos synchronization.

\end{abstract}

\end{frontmatter}
\bibliographystyle{elsart-num}
\sloppy

\section{Introduction}
The feasibility of enslaving two chaotic systems \cite{pecora90}
opened the possibility of using the signals generated by chaotic
systems as carriers for analog and digital communications and soon
aroused great interest as a potential means for secure
communications \cite{Yang04}. It is assumed in the literature that
chaotic modulation is an adequate means for secure transmission,
because chaotic maps present some properties, such as sensitive
dependence on parameters and initial conditions, ergodicity,
mixing, and dense periodic points, that make them similar to
pseudo random noise \cite{devaney92}, which has been used
traditionally as a masking signal for cryptographic purposes.

For over a decade a number of secure communication systems have
been proposed in which the plaintext message signal $m(t)$ was
concealed into the chaotic signal by simply adding it to a system
variable $u(t)$ of the sender chaotic generator
\cite{Boutayeb02,Memon03,Bowong04}; the receiver had to
synchronize with the sender to regenerate the chaotic signal
$\tilde{u}(t)$ and thus recover the message $m(t)$. This
uncomplicated scheme is usually broken by setting apart $u(t)$ and
$m(t)$ signals using elemental high pass filtering
\cite{Alvarez04f,Alvarez04g,Alvarez05b}, or by directly estimating
the chaotic signal $u(t)$ via Short's NLD method
\cite{Short94,Short97}.

To avoid this weakness a more elaborated mixing procedure was
employed in some recently proposed chaotic cryptosystems: a
two-channel transmission technique was used, where an unmodified
chaotic system variable was transmitted using the first channel,
while a second channel conveyed a signal that was a complicated
non-linear combination of the plaintext and another system
variable, from which it was impossible to retrieve both
separately. The first channel served as synchronizing signal for
the chaotic system receiver, then the remaining chaotic system
variables were generated and employed at the receiver end to
retrieve the plaintext from the second channel signal, using the
same system parameters values at sender and receiver
\cite{Jiang02,Wang04,Li04}.

In the vast majority of chaotic cryptosystems the security relies on
the secrecy of the system parameters, which play the role of secret
key, hence the determination of the system parameters from the
chaotic ciphertext is equivalent to breaking the system.

The contribution of this work is double. First, a novel
determination method of the unknown parameters of a Lorenz system,
when the waveform of one of its variables is known, is presented
in Sec.~\ref{parameter}. Then, in Secs.~\ref{twochannel} and
\ref{twochannel2}, it is shown how this method can be applied to
break some two-channel cryptosystems that use the Lorenz chaotic
system. Finally, Sec.~\ref{Sec:Conclusion} concludes the paper.

\section{Parameter determination of the Lorenz system}
\label{parameter}

Since 1963 the Lorenz system \cite{Lorenz63} has been a paradigm
for chaos. Consequently, it has been  predominantly used in the
design of chaotic cryptosystems. It is defined by the following
equations:
\begin{align}
  \dot {x}  &= \sigma(y-x),\nonumber\\
  \dot {y}  &=rx - y -xz,\label{sender}\\
  \dot {z}  &=xy - b z.\nonumber
\end{align}
where $\sigma$, $r$ and $b$ are fixed parameters.

The proposed approach to the problem of Lorenz system parameter
determination is based on a homogeneous driving synchronization
mechanism \cite{Pecora91} between a drive Lorenz system and a
response subsystem that is a partial duplicate of the drive system
reduced to only two variables, driven by the third variable.

Projective synchronization (PS) is an interesting phenomena firstly
described by Mainieri and Rehacek \cite{Mainieri99}. It consists of
the synchronization of two partially linear coupled chaotic systems,
sender and receiver, in which the amplitude of the slave system is a
scalar multiple, called scaling factor, of that of the sender system
in the phase space. The original study was restricted to
three-dimensional partially linear systems. Xu and Li \cite{Xu02b}
showed that PS could be extended to general classes of chaotic
systems without partial linearity, by means of the feedback control
of the slave system.

The response system is defined by the following equations, in
which variable z(t) is used as driving signal:
\begin{align}
  \dot {x}_r  &= \sigma^*(y_r-x_r),\nonumber\\
  \dot {y}_r  &= r^* x_r - y_r -x_r z,\label{intruder}
\end{align}
where $\sigma^*$ and $r^*$ are fixed parameters and the drive
variable is $z$.

As was shown in \cite[$\S$III]{Pecora91} this drive-response
configuration has two conditional Lyapunov exponents, the first
one is fairly negative while the second one is of small positive
value, thus leading to a slightly unstable system. The consequence
is that if the parameters of drive and response systems are
identical, then the drive and response variables will become
identical (for complete synchronization) or differ only in an
scaling factor (for projective synchronization), that depends on
the initial conditions of the drive and response systems. However,
if the parameters are not exactly equal, then the drive and
response variables will be completely different.

When the drive and response systems parameters are equal, the
variable $x_r(t)$ will be easily recognizable as the familiar
waveform of a Lorenz system, by a supposed human skilled observer.
But if drive and response systems parameters are different, the
waveforms generated by the response system will be a nonsense mesh
some seconds after the beginning of driving, due to the sensitive
dependence of chaotic systems on parameter values. This phenomenon
could be interpreted by the observer as the consequence of a wrong
parameter guessing.

This work describes a criterion, based on the study of some
geometric properties of the Lorenz system variables waveforms, to
automatically decide if the response system parameters coincide
with the drive system parameters or not, by means of the analysis
of the $x_r(t)$ waveform of the response system.

This method of recovering the unknown system parameters is
applicable in the case of cryptosystems that use the variable $z(t)$
as the driving signal like \cite{Wang04,Li04}. But it is not
applicable to other two-channel cryptosystems driven by $x(t)$ or
$y(t)$, like \cite{Jiang02}, because in those cases both Lyapunov
exponents are negative and the drive-response configuration is
stable, in despite of being the drive and response parameters
moderately different. In those cryptosystems another efficient
method of revealing the Lorenz system parameters described by
Stojanovski \emph{et al.} is aplicable \cite{Stojanovski96}.

To minimize the computer workload as much as possible, the
parameter search space is previously reduced to a narrow range by
means of a simple measure upon the $z(t)$ waveform. Then, all the
unknown parameter values are determined with the desired accuracy.

\subsection{Lorenz attractor's geometrical properties}

According to \cite{Lorenz63} the Lorenz system has three fixed
points. For $0<r<1$ the origin of coordinates is a globally stable
fixed point; for $1\leq r<r_c$ the origin becomes unstable giving
rise to two other stable twin points $C^+$ and $C^-$, of
coordinates $C^\pm=(\pm\sqrt{b(r-1)},\pm\sqrt{b(r-1)},(r-1))$,
being $r_c$ a critical value defined as:
\begin{equation}\label{eq:critical}
r_c=\frac{\sigma(\sigma+b+3)}{\sigma-b-1}.
\end{equation}
When $r$ exceeds the critical value $r_c$, the system becomes
unstable, and its behavior is chaotic.

\begin{figure}[t ]
\begin {center}
\psfrag{x}{$x$}  \psfrag{z}{$z$}
\begin{overpic}[scale=1]{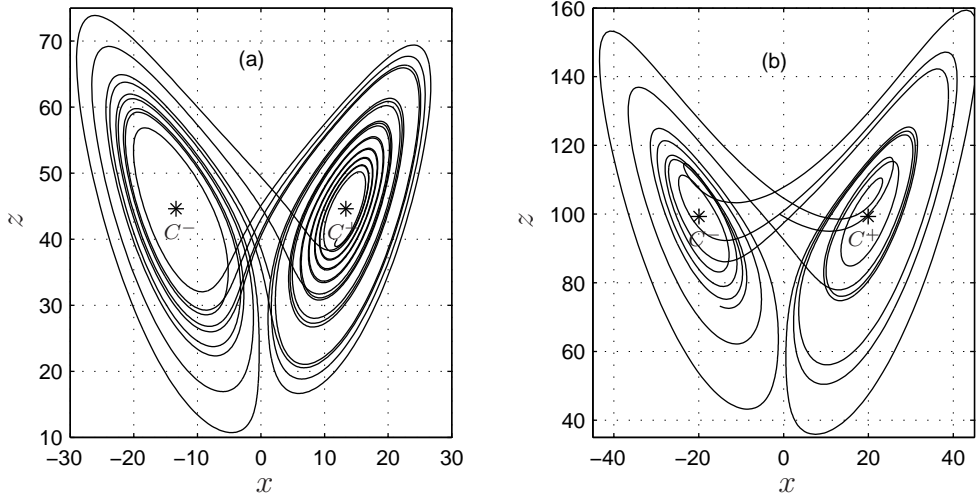}
    \put(33.5, 26){\scriptsize$C^+$}
    \put(17, 26){\scriptsize$C^-$}
    \put(86.1, 25.3){\scriptsize$C^+$}
    \put(70,   25.3){\scriptsize$C^-$}
\end{overpic}
\caption{\label{fig:mari2} Lorenz chaotic attractor: (a)
parameters $r=45.6$, $\sigma=16$ and $b=4$; (b) parameters
$r=100.3$, $\sigma=16$ and $b=4$, showing irregular cycles that
not surround the fixed points. The position of the fixed points
$C^+$ and $C^-$ is indicated by asterisks.}
\end{center}
\end{figure}

The Fig.~\ref{fig:mari2}(a) shows the well known double scroll
Lorenz attractor formed by the projection on the $x-z$ plane, in
the phase space, of a trajectory portion extending along 10 sec,
where the parameters are $r=45.6$, $\sigma=16$ and $b=4$, the
initial conditions are $x_0=13.3566$, $y_0=13$ and $z_0=44.6$, the
fixed points $C^+$ and $C^-$ are indicated by asterisks.

It is a well known fact that the Lorenz attractor trajectory
follows two loops, in the vicinity of the fixed points $C^+$ and
$C^-$, with a spiral-like shape of steadily growing amplitude,
jumping from one to the other, at irregular intervals, in a
random-like manner though actually deterministic \cite{Lorenz63}.
The trajectory always jumps from a cycle of relative high
amplitude to another on the opposite  loop generally of smaller
amplitude. The spiraling trajectory may pass arbitrarily near to
the fixed points, but never reach them while in chaotic regime.

\textit{Definition 1.} The portions of the attractor trajectory that
consist of a revolution of $360 ^{\circ}$ beginning after a change
of sign of $x$ and $y$ are \textit{irregular cycles}. The portions
of the trajectory that constitute a complete spiral revolution of
$360 ^{\circ}$ and do not begin after a change of sign of $x$ and
$y$ are \textit{regular cycles}.

\textit{Remark 1.} Regular cycles always surround the fixed points
$C^+$ or $C^-$, taking them as centers of a growing spiral.

\textit{Remark 2.} Irregular cycles usually surround the fixed
points $C^+$ or $C^-$; but sometimes may not surround them,
instead the trajectory may pass slightly above them in the $x-z$
plane. This phenomenon is illustrated in the
Fig.~\ref{fig:mari2}(b), with system parameters $r=100.3$,
$\sigma=16$ and $b=4$ and initial conditions $x_0=-1$, $y_0=35.24$
and $z_0=100$.

\textit{Definition 2.} The \textit{attractor eyes} are constituted
by the two neighborhood regions around the fixed points that are not
filled with regular cycles. The eye centres are the fixed points
$C^+$ or $C^-$.

\textit{Definition 3.} The \textit{eye aperture} $x_a$ and $z_a$
of the variables $x$ and $z$, for a particular time period, is the
smallest distance between the maxima and minima of $|x(t)|$ and
$z(t)$, respectively, of the regular cycles, measured along this
time period.

\begin{figure}[h]
\begin{center}
\psfrag{x}{$|x|$}  \psfrag{z}{$z$}
\begin{overpic}[scale=1]{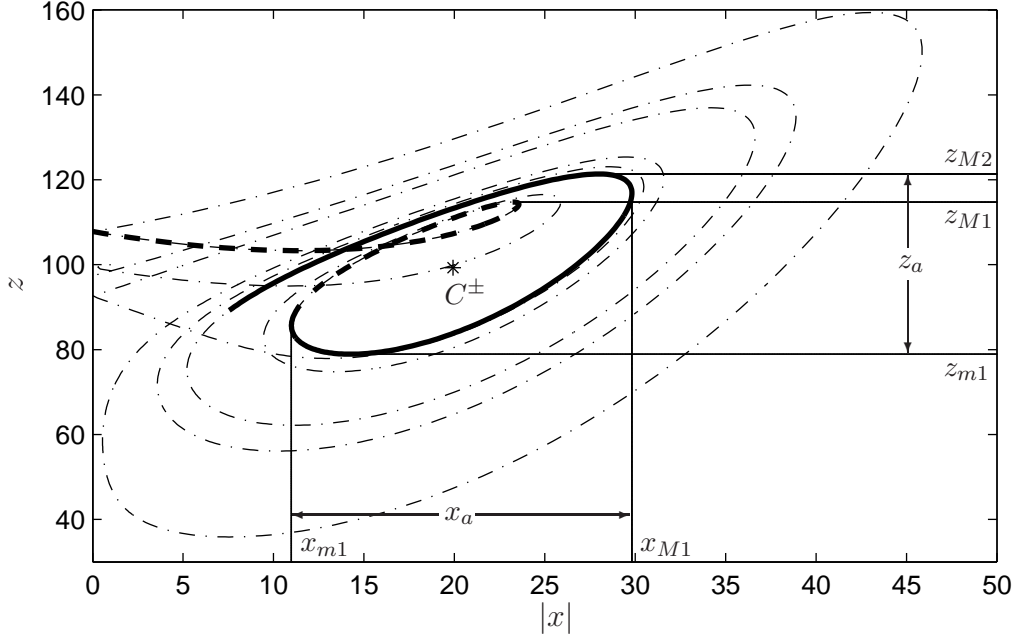}
    \put(44,32.5){\footnotesize$C^\pm$}
    \put(44,11.5){\footnotesize$x_a$}
    \put(47,12){\vector(1,0){15}}
    \put(43.4,12){\vector(-1,0){14.2}}
    \put(30,8.4){\footnotesize$x_{m1}$}
    \put(63,8.4){\footnotesize$x_{M1}$}
    \put(88,36){\footnotesize$z_a$}
    \put(89,38){\vector(0,1){7}}
    \put(89,34.7){\vector(0,-1){7}}
    \put(92.5,40.3){\footnotesize$z_{M1}$}
    \put(92.5,46.3){\footnotesize$z_{M2}$}
    \put(92.5,25.5){\footnotesize$z_{m1}$}
\end{overpic}
\caption{First 2.25 s of a version of the Lorenz attractor of
Fig.~\ref{fig:mari2}(b), folded around the $z$ axis. The solid
thick line trajectory portion is the regular cycle closest to the
fixed points $C^\pm$. The dashed thick line trajectory portion is
the preceeding irregular cycle.} \label{fig:todo}
\end{center}
\end{figure}

Figure~\ref{fig:todo} illustrates the first 2.25 s of another
version of the Lorenz attractor of Fig.~\ref{fig:mari2}(b), folded
around the $z$ axis and formed by the projection on the $x-z$
plane, in the phase space, of a trajectory portion of $z(t)$ and
$|x(t)|$. The trajectory portion drawn with solid thick line is
the regular cycle closest to the fixed points $C^\pm$, from which
the eye aperture of $x_a$ and $z_a$ can be determined. The
trajectory portion drawn with dashed thick line belongs to the
preceding irregular cycle.

\subsection{Reduction of the parameters search space}
\label{reduction}

The geometrical properties of Lorenz system allows for a previous
reduction of the search space of the $r$ parameter, before carrying
out the accurate parameter determination, taking advantage of the
relation of the system parameter $r$ with the coordinates
$z_{C^+}=z_{C^-}=r-1$ of the fixed points $C^+$ and $C^-$ and
Eq.~(\ref{eq:critical}). The estimated value $z^*_{C^\pm}$ of the
fixed points coordinates $z_{C^+}=z_{C^-}$ was calculated from the
variable $z(t)$ using following algorithm:

\begin{enumerate}
    \item compile a list of all the relative maxima and minima of $z(t)$,
    \item exclude all the minima belonging to an irregular cycle from the list,
    \item retain the biggest relative minimum $z_{m1}$, among the remaining list elements,
    \item select the two maxima $z_{M1}$, $z_{M2}$ immediately preceding and following
    $z_{m1}$, respectively,
    \item calculate the spiral centre as $z^*_{C^\pm}=(\frac{1}{3}z_{M1}+\frac{2}{3}~z_{M2}+z_{m1})/2$.
\end{enumerate}

There is no need to find a rule of growing for the spiral radius,
since the optimal values of the two weights of $z_{M1}$ and
$z_{M2}$, in the preceding $z^*_{C^\pm}$ formula, can be determined
experimentally.

The minima of the irregular cycles were discarded because they are
inappropriate for the fixed point's $z$ coordinate calculation, due
to the fact that irregular cycles may not take the fixed points as
centres. Those cycles are very easy to detect from the $z(t)$
waveform: they are the first minima that comes after a previous
minimum of smaller value.

\begin{figure}
\begin{center}
\begin{overpic}[scale=1]{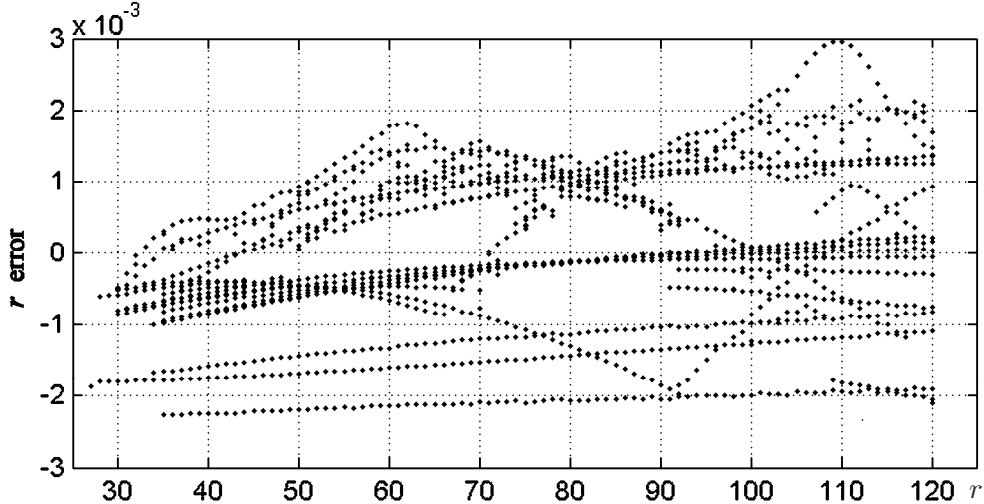}
     \put(99,0.6){\footnotesize$r$}
\end{overpic}
\caption{Parameter $r$ estimation error, when calculated from the
fixed point coordinate $z^*_{C^\pm}$, for different combinations
of system parameters $\sigma$ and $b$.} \label{fig:r-error}
\end{center}
\end{figure}

Figure~\ref{fig:r-error} illustrates the relative error when the
value of $r$ is estimated as $r^*=z^*_{C^\pm}+1$, for values of
$r^*$ ranging from the critical value $r^*=r_c$ to $r^*=120$, in
increments of $\Delta r^*=1$, for 15 different combinations of
system parameters, $\sigma=(6,10,13,16,20)$ and $b=(2,8/3,4)$; the
analyzed time was 200 s of the $z(t)$ waveform. As can be seen, the
maximum relative error spans from $-0.23\%$ to $+0.3\%$. In this
way, when trying to guess the value of $r$ from the waveform of
$z(t)$, the effective search space may be reduced to a narrow margin
of less than $0.6\%$ of the computed value $r^*=z^*_{C^\pm}+1$.

The presence of moderate noise added to the $z(t)$ waveform did
not affect the precision of the measure. Some tests were carried
adding either white gaussian noise or sinusoidal signals, of a
level 30 db below $z(t)$. The resultant relative error in the
guess of $r^*$ was still inferior to $\pm0.2\%$, for $\sigma=16$
and $b=4$. But for not so moderated values of added noise the
increase of relative error was noticeable; i.e. when the noise
reached a value of 20 db below $z(t)$ the relative error raised to
about $\pm1\%$.

The search space of $\sigma^*$ can also be delimited. Assuming
that $r>r_c$, $b \geq 0$ and $\sigma >0$, it follows from Eq.~
(\ref{eq:critical}) that:
\begin{equation}
\label{eq:sigma}
0>\sigma^2+(b+3-r)\sigma +r(b+1)>\sigma^2+(3-r)\sigma,
\end{equation}
that yields a very conservative margin of $0<\sigma<r-3$.

\subsection{Accurate parameter determination}
\label{accurate}

Once the search space of the parameters is fixed, a homogeneous
driving synchronization based procedure can be implemented to
determine the approximate values $r^*$  and $\sigma^*$ with any
desired accuracy. For this purpose, the response system described
by Eq.~(\ref{intruder}) was used.

When the synchronizing signal is fed to the response and the
parameters of both systems agree, i.e. $r^*=r$ and
$\sigma^*=\sigma$, the variables $x_r$ and $y_r$ follow the drive
signals $x$ and $y$ with a scale factor that depends on the
initial conditions. If the parameters of both systems do not
agree, i.e. $r^* \neq r$ and/or $\sigma^* \neq \sigma$, the
variables waveforms of drive and response systems will differ
absolutely, even if the initial conditions are the same. After a
few system iterations, all waveforms generated with different
parameters values are nearly alike, but as the number of
iterations grow, the waveforms generated with different parameter
values begin to diverge, due to the conditional positive Lyapunov
exponent of the drive-response configuration. For large number of
iterations, even the smallest difference in parameters values
leads to a serious disagreement of drive and response waveforms.

\begin{figure}[t]
\begin{center}
\psfrag{z}{$z$} \psfrag{xi}{$x_r$}
\includegraphics{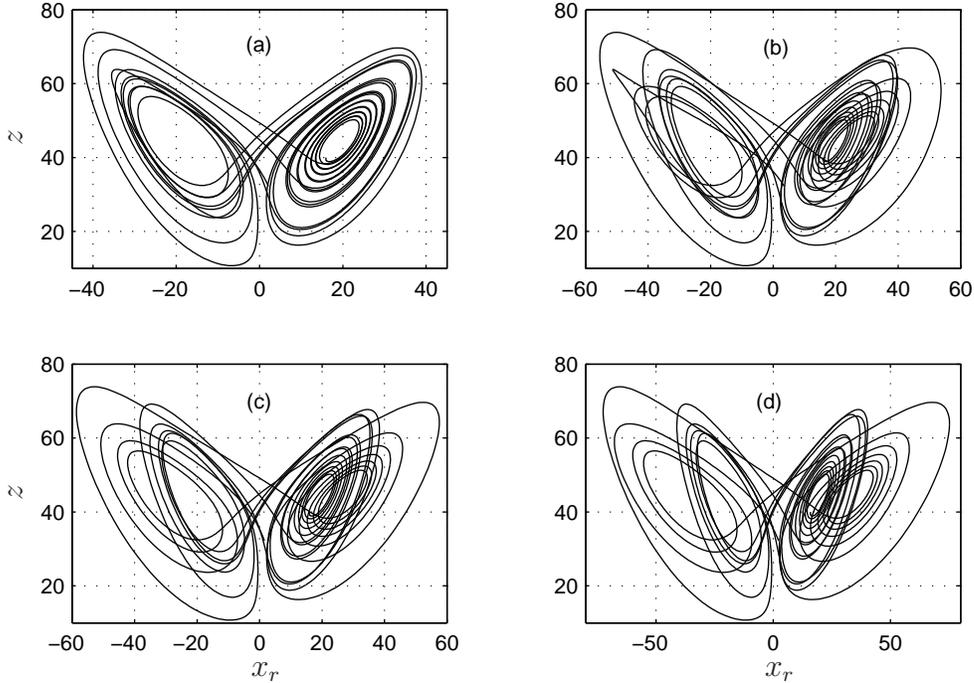}
\caption{Lorenz attractor formed by the projection on the $x_r-z$
plane, in all cases the drive parameter are the same: $\sigma=16$
and $r=45.6$; but response parameter values are different: (a)
$\sigma^*=\sigma$, $r^*=r$; (b) $\sigma^*=\sigma$, $r^*=45.61$;
(c) $\sigma^*=15.65$, $r^*=r$; (d) $\sigma^*=15.65$, $r^*=45.61$.}
\label{fig:mari4}
\end{center}
\end{figure}

Figure~\ref{fig:mari4} shows the double scroll Lorenz attractor
formed by the projection on the $x_r-z$ plane when four possible
cases of parameter coincidence are considered. In
Fig.~\ref{fig:mari4}(a) both parameters of drive and response
systems are equal. It can be seen that the attractor is similar to
the illustrated in Fig.~\ref{fig:mari2}(a), being the difference
the disagreement in the horizontal scale due to different initial
conditions, it can be also observed that the attractor eye is
quite open. In Fig.~\ref{fig:mari4}(b) one parameter coincides,
but the other differs: $\sigma=\sigma^*=16$,  $r=45.6$ and
$r^*=45.61$, it can be seen that eye aperture has diminished
considerably with respect to the former case. In
Fig.~\ref{fig:mari4}(c) the coinciding parameter is $r=r^*=45.6$,
the differing one is $\sigma=16$ and $\sigma^*=15.65$, it can be
seen that the eye aperture has diminished even more. Finally, in
Fig.~\ref{fig:mari4}(d) both parameters differ $r=45.6$,
$r^*=45.61$, $\sigma=16$ and $\sigma^*=15.65$, it can be seen that
the eye is completely closed, i.e. the eye $x$-aperture $x_a$ is
negative. Similar measures were carried out for a great variety of
drive parameter values with identical results. When the
differences between the true parameter values and the guessed
values $r-r^*$ and $\sigma-\sigma^*$ are big, the eye aperture
closes after very few cycles, but for progressively diminishing
differences between parameters the number of cycles needed to
obtain a closing eye are increasing.

The value of the eye $x$-aperture $x_a$ of the variable $x_r(t)$
was computed for many sets of parameters values. It was found in
all cases that its maximum value was reached when $r^*=r$ and
$\sigma^*=\sigma$. For these parameter values the variables $x$
and $x_r$ are completely synchronous but differ only in a
proportionality factor. Hence the maximum eye aperture is an
excellent numerical criterium for evaluating the synchronism
between drive and response systems.

The eye $x$-aperture $x_a$ of the variable $x_r(t)$, was
calculated with the following algorithm:
\begin{enumerate}
    \item compile a list of all relative maxima and minima of $\textrm{abs}(x_r(t))$,
    \item exclude all the maxima belonging to an irregular cycle from the list,
    \item retain the smallest relative maximum $x_{M_1}$, among the remaining maxima,
    \item select the biggest minimum $x_{m1}$, among all the minima,
    \item calculate the eye aperture as $x_a=x_{M1}-x_{m1}$.
\end{enumerate}

\section{Application to cryptanalysis of a multiplexed two-channel projective
synchronization cryptosystem}\label{twochannel}

After their research in PS, Xu and Li proposed a secure
communication scheme based on PS chaotic masking \cite{Li04}, that
was shown to be breakable by filtering and by generalized
synchronization using the feedback of the plaintext recovery error
\cite{Alvarez05a}.

In a recent article Wang and Bu \cite{Wang04} proposed a new
encryption scheme also based on PS. Following \cite{Mainieri99},
the state vector of a partially linear system of ordinary
differential equations was broken in two parts $(\textbf{u},z)$.
The equation for $z(t)$ was nonlinearly related to the other
variables, while the equation for the rate of change of the vector
$\textbf{u}$ was linearly related to $\textbf{u}$ through a matrix
$M$ that may depend on the variable $z(t)$. It was employed a
sender system $(\textbf{u}_s,z)$, a receiver system
$(\textbf{u}_r,z)$, and an auxiliary system $(\textbf{u}_c,z)$
defined as:
\begin{align}
  \dot{\textbf{u}}_s &=M(z)\cdot\textbf{u}_s,\quad \dot{z}=
  f(\textbf{u}_s,z),\nonumber\\
  \dot{\textbf{u}}_r &=M(z)\cdot\textbf{u}_r,\label{all}\\
  \dot{\textbf{u}}_c &=M(z)\cdot\textbf{u}_c,\nonumber
\end{align}
where $\textbf{u}_s=(x_s,y_s)$, $\textbf{u}_r=(x_r,y_r)$, and
$\textbf{u}_c=(x_c,y_c)$. When PS takes place
$\lim_{t\rightarrow\infty}\|\textbf{u}_s-\alpha\textbf{u}_r\|=0$,
being $\alpha$ a constant depending on the initial conditions of
$\textbf{u}_r(0)$ and $\textbf{u}_s(0)$.

The ciphertext $s(t)$ was defined as a composition in function of
time of the shared scalar variable $z(t)$ and the scalar variable
$x_s(t)$, described as:
\begin{equation}
s(t) = \left\{ {\begin{array}{*{20}l}
   {x_s (t),\quad n\Delta t \le t \le n\Delta t + \delta t,}  \\
   {z(t),\quad n\Delta t + \delta t < t \le (n+1)\Delta t,}  \\
\end{array}} \right.
\end{equation}
being $n=(0,1,2,\ldots)$, while $\Delta t$ and $\delta t$ are two
time intervals so that $\delta t\ll \Delta t$.

The ciphertext plays the double roles of the driving signal for
chaos synchronization between the sender and receiver, by means of
$z(t)$, and the message carrier through $x_s(t)$.

It is supposed that the plaintext message $i(t)$ was previously
discretized in time, in the form of a string of bits or a string
of samples, $i_n$. In the first case, the bits are coded as $+1$
or $-1$. In the second case, the analog signal is sampled at a
rate of $1/\varepsilon$ samples per second, where $\varepsilon$ is
the sampling period.

The encryption of a plaintext $i(t)$ was achieved as follows: at the
beginning of each time interval $\Delta t$, during a much shorter
time interval $\delta t$, the sender system vector $\textbf{u}$ is
forcibly modified in the following way:
\begin{equation}
\textbf{u}_s(t_n)=i_n \textbf{u}_c(t_n),
\end{equation}
and at the end of the time interval $\delta t$ the entire system was
let freely evolving until the beginning of the next time period
$\Delta t$.

\begin{figure}[tb]
\begin{center}
\psfrag{x(t)}{$x_s(t)$}\psfrag{s(t)}{$s(t)$}\psfrag{t}{$t$}
\includegraphics[scale=1]{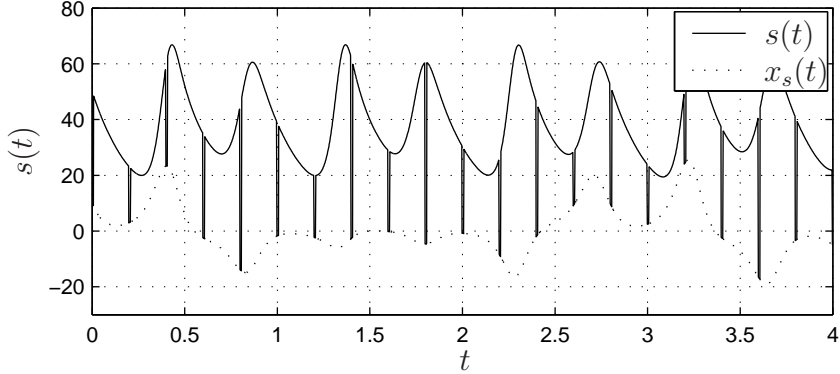}
\caption{Ciphertext $s(t)$, for $\Delta t=0.2$ and $\delta
t=0.01$~(solid line) and scalar variable $x_s(t)$ (dotted line).}
\label{fig:ciphertext}
\end{center}
\end{figure}

Figure~\ref{fig:ciphertext} illustrates the waveform of the
ciphertext. It can be seen that $s(t)$ is a discontinuous signal
that agrees most of the time with the function $z(t)$, but jumps
to the value of $x_s(t)$ during a small time interval $\delta t$
every $\Delta t$ seconds.

The function $z(t)$ was easily recovered, at the receiver end, by
filtering out the spikes. The final signal distortion is
negligible due to the short spike time length $\delta t$ related
to their repetition period $\Delta t$.

To recover the plaintext, instead of using the signal $x_s(t)$,
which is not available at the receiver end, the average value of
the spikes $\bar{x}_s(t)$ for $n\Delta t \le t \le n\Delta t +
\delta t$ was employed. Thanks again to the fact that $\delta t
\ll \Delta t$, it can be considered that $\bar{x}_s(t)$ is a good
approximation of $x_s(t)$.

The recovered plaintext $i'_n(t)$ at the receiver end was
calculated as:
\begin{equation}
i'_n(t)=
\frac{\bar{x}_s(t_n)}{x_r(t_n)}=\frac{\bar{y}_s(t_n)}{y_r(t_n)}
\end{equation}

If the initial conditions of the auxiliary system and the receiver
system are identical, the original plaintext and the retrieved
plaintext will agree: $i'_n(t)=i_n(t)$. However, if the initial
conditions are different, the retrieved plaintext will be not
equal, but proportional, to the original plaintext:
$i'_n(t)=c\,i_n(t)$. Due to PS between the sender and receiver,
here \textit{c} is a constant.

For practical purposes, the present system is a two-channel
communication system, with the particularity that both channels,
one continuos and another sampled, are transmitted in a
multiplexed way and separated at the receiving end.

In \cite[\S3]{Wang04} an example was presented using similar
sender-receiver circuits to the ones described in \cite{Cuomo93b},
based on the Lorenz system, see Eq.~(\ref{sender}), with parameter
values:
\begin{equation}
\sigma=16.0,~ r=45.6,~ b=4.0,~ \Delta t=0.2,~ \delta t=0.01,~
\varepsilon=0.001.  \label{param1}
\end{equation}
It was shown that an absolute error of $\Delta r^*=0.001$ in the
value of the receiver parameter $r^*$ leads to a plaintext
recovery failure, and it was asserted that a similar deviation in
the receiver parameter $\sigma^*$ value has the same effect.
Hence, although not clearly stated by the authors of
\cite{Wang04}, it can be estimated that in this cryptosystem the
parameter values play the role of secret key. In every
cryptosystem, the key should be completely specified
\cite{AlvarezLi06}.

The authors of \cite{Wang04} claimed that this method has some
remarkable advantages over other chaos-based secure communication
schemes, because it is not possible to extract the plaintext
directly from the ciphertext by means of an error function attack,
due to the system high sensitivity to the parameter values.
Moreover, conventional return map attacks exploiting the
perturbation of the sender dynamics are also avoided, because the
modulation procedure only affects the initial values of the
trajectories in the phase space.

\subsection{System parameters recovery procedure} \label{s:recovery}

In the system proposed in \cite{Wang04}, the variable $z(t)$ was
extracted at the receiver end from the ciphertext $s(t)$ and used
to achieve the receiver synchronization. This fact allows mounting
an attack against the system parameters, whose values can be
accurately determined.

In our simulation, the same sender used in \cite{Wang04} was
employed as a drive system, described by Eq.~(\ref{sender}). As
the intruder's receiver the response system described by
Eq.~(\ref{intruder}) was used. The same sender parameters of the
authors' example were used. The initial conditions of the sender
were arbitrarily chosen as $x_s(0)=40$, $y_s(0)=40$, $z(0)=40$,
because in \cite{Wang04} there was no details about them. The
initial conditions of the intruder's response system were
arbitrarily chosen as $x_r(0)=70$, $y_r(0)=7$.

The adequate searching ranges for the parameters $r^*$ and
$\sigma^*$ were determined as follows: applying the algorithm
described in the Section~\ref{reduction} to 200 s of the $z(t)$
waveform, it was found that the fixed point $z$ coordinate was
$z^*_{C^\pm}=44.5943$, that corresponds to $r^*=45.5943$ (very
close to the true value $r=45.6$); hence a practical search range
of $r^*$ going from $r^*=45.50$ to $r^*=45.70$ was selected,
equivalent to an error allowance ranging from $-0.23\%$ to
$+0.2\%$, compliant with Fig.~\ref{fig:r-error}. The search space
of $\sigma^*$, according to Eq.~(\ref{eq:sigma}), should be
comprised in the range $0 < \sigma^* <42.70$.

\begin{figure}[t]
\begin{center}
\psfrag{Eye x aperture}{\scriptsize Eye $x$ aperture}
\includegraphics{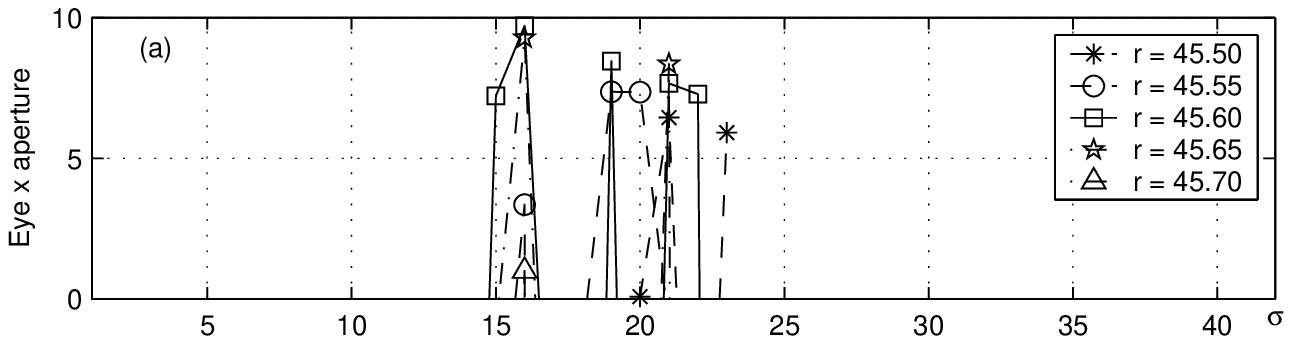}
\includegraphics{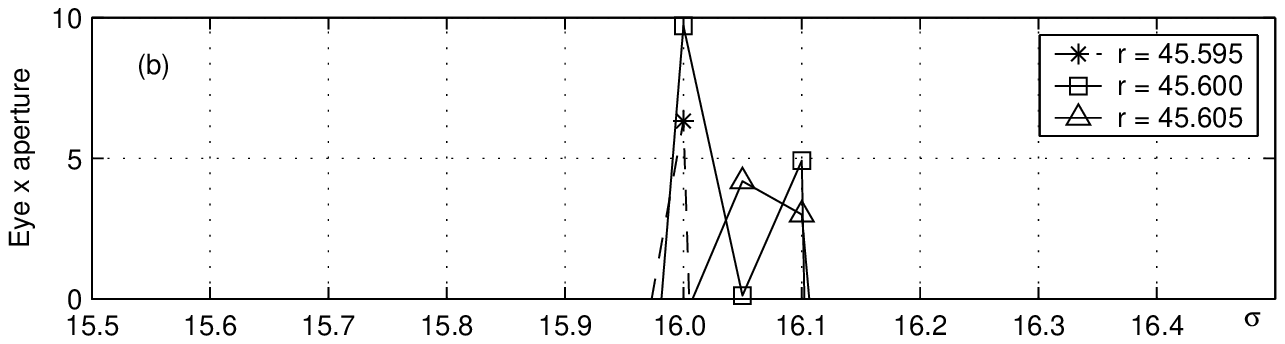}
\includegraphics{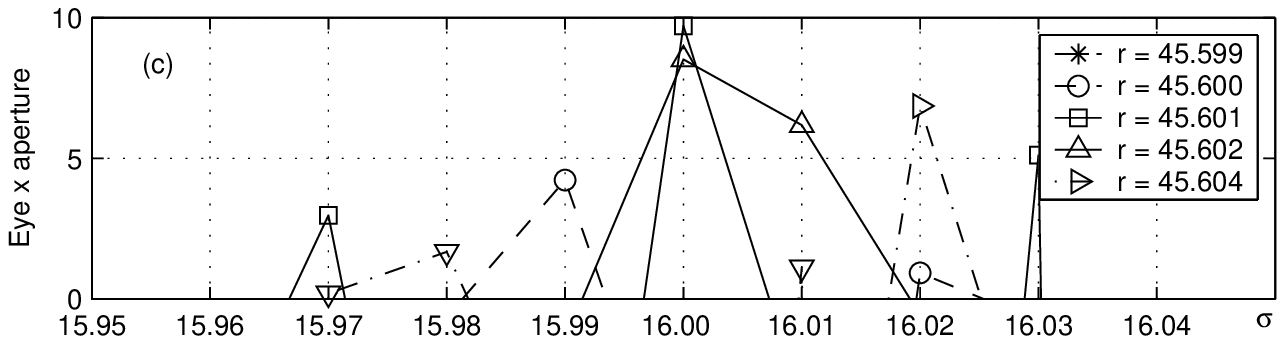}
\includegraphics{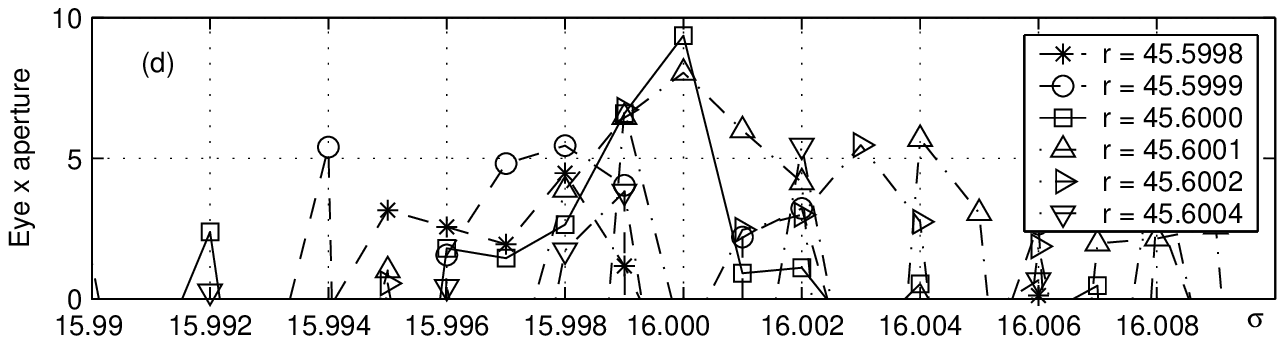}
\begin{overpic}[scale=1]{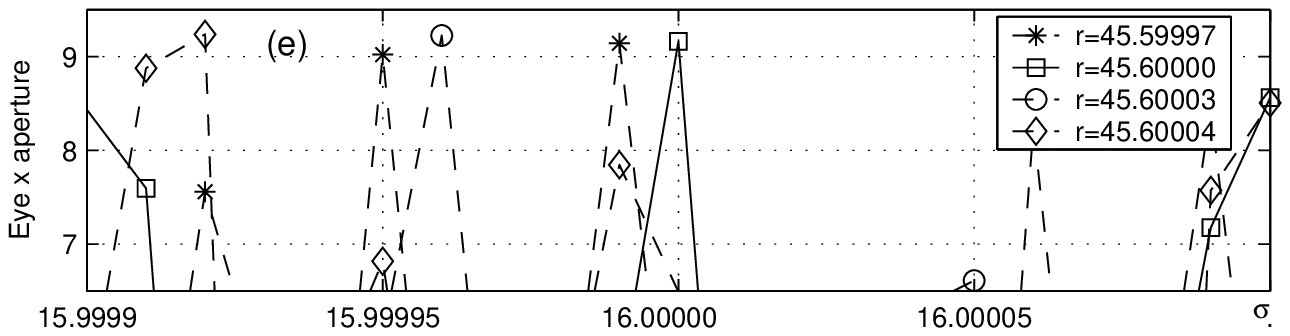}
     \put(9,23){\vector(1,0){5}}
     \put(27,23){\vector(1,0){5}}
     \put(41,22.3){\vector(1,0){5}}
     \put(59,22.3){\vector(-1,0){5}}
\end{overpic}
\caption{\label{fig:s} Intruder receiver eye aperture $x_r$ for
various measure periods: (a) 25 s; (b) 80 s; (c) 250 s; (d) 800 s;
(e) 800 s}
\end{center}
\end{figure}

Figure \ref{fig:s} illustrates the $r^*$ and $\sigma^*$
determination method using the procedure described in
S.~\ref{accurate}, that is accomplished in five steps. In the
first step, the eye aperture of the receiver $x_r$ variable was
measured along a period of 25 s, equivalent to 55 periods of
$z(t)$. The measure was made for each of the 210 different sets of
parameter values obtained varying $r^*$ from $r^*=45.50$ to
$r^*=45.70$, in increments of $\Delta r^*=0.05$, and $\sigma^*$
from $\sigma^*=1$ to $\sigma^*=42$, in increments of $\Delta
\sigma^*=1$. The result is illustrated in Fig.~\ref{fig:s}~(a). It
can be seen that for most combinations of parameter values the
aperture is negative, i.e. the corresponding parameter values are
far from the right value; the best values for $\sigma^*$ are
comprised between $\sigma^*=15.5$ and $\sigma^*=16.5$, while the
best values for $r^*$ are comprised between $r^*=45.55$ and
$r^*=45.65$. Those values are taken as the search limits for the
next step. The mesures were done, in the second, third and fourth
steps, during periods of 80 s, 250 s and 800 s respectively, the
results are depicted in Figs.~\ref{fig:s}(b), \ref{fig:s}(c) and
\ref{fig:s}(d).

If the available ciphertext was unlimited, the next measure step
could be done over a period longer than 800 s until the desired
parameter precision could be reached. But let us suppose that
there is no more than 800 s of available ciphertext. In that case,
the only choice is to constrict the search space around the last
best result obtained, with a growing resolution, until a situation
is reached in which it becomes impossible to decide which is the
best parameter value. The Fig.~\ref{fig:s}~(e) illustrates this
situation, it was obtained keeping the last measure period of 800
s, but narrowing the search space around the last best result
obtained. It can be seen that the discrimination limit of the
identification method was reached for that period of measure,
because multiple peaks gave approximately the same eye aperture of
$x_a\approx 9.2$. The four peaks of greater amplitude suggest four
sets of equally plausible potential candidates of response system
parameter sets, one of them is the right one $r^*_0=r=45.60000$,
$\sigma^*_0=\sigma=16.00000$, the other three are slightly
inexact, they differ in the seventh significative digit from the
right value: $r^*_1=45.59997$, $\sigma^*_1=15.99999$;
$r^*_2=45.60003$, $\sigma^*_2=15.99996$ and $r^*_3=45.60004$,
$\sigma^*_3=15.99992$.

\begin{figure}
\begin{center}
\psfrag{x(t)}{$x_s(t)$} \psfrag{xi(t)}{$x_r(t)$}
\begin{overpic}[scale=1]{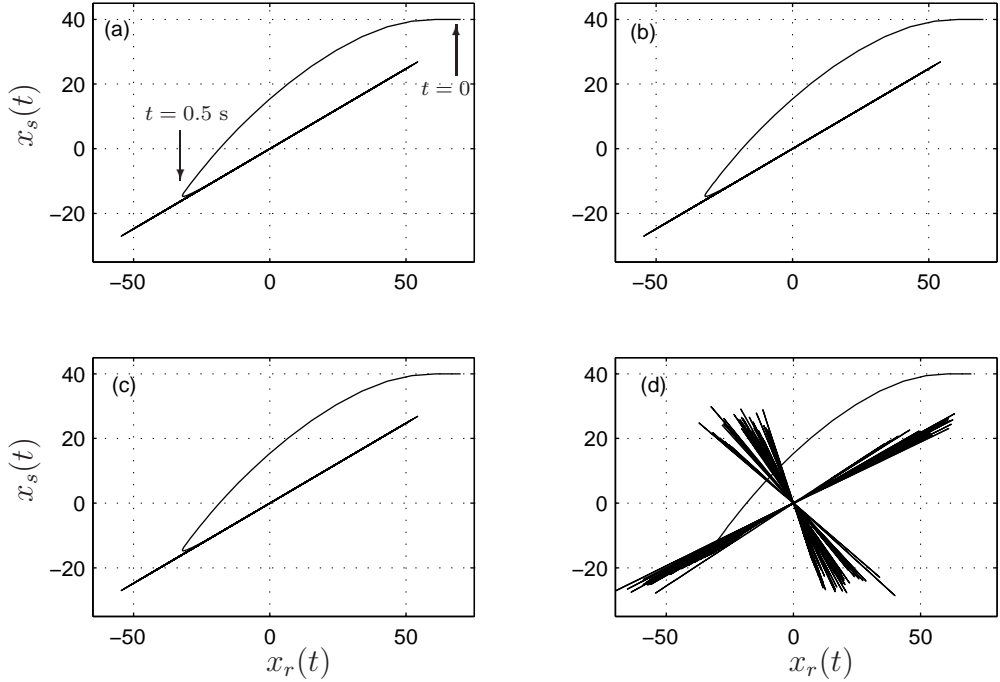}
     \put(45,   60.5) {\vector(0,1){5}}
     \put(41.5, 58.5) {\scriptsize$t=0$}
     \put(17.5, 55)   {\vector(0,-1){5}}
     \put(14,   56)   {\scriptsize$t=0.5$~s}
\end{overpic}
\caption{\label{fig:error} First 800 s of the intruder receiver
phase portrait, for various sets of response system parameters:
(a) $r^*=45.59997$, $\sigma^*=15.99999$; (b) $r^*=45.60003$,
$\sigma^*=15.99996$; (c) $r^*=45.60004$, $\sigma^*=15.99992$; (d)
$r^*=45.601$, $\sigma^*=15.999$.}
\end{center}
\end{figure}

The Figs.~\ref{fig:error}(a), \ref{fig:error}(b) and
\ref{fig:error}(c) illustrate the 800 first seconds of the
waveform of $x_r(t)$ plotted against $x(t)$, for the tree inexact
system parameter sets. It can be seen that the $x_r(t)$ and $x(t)$
waveforms are perfectly correlated in all the three cases despite
of the parameter values little inexactitude. The different initial
conditions are the cause of the initial transitory, that lasts
only 0.5 s and of the different scale amplitudes of the waveforms.
This means that any of the four potential candidates of response
system parameter sets may be used indistinctly to generate the
$x_r(t)$ waveform without noticeable error, for the limited time
period that was considered for their determination.

For practical purposes, a limited precision in the determination
of the parameters is not a shortcoming, because the degree of
coincidence of the eye apertures $x_{a1}$ and of $x_{a2}$ of two
waveforms of $x_{i1}(t)$ and $x_{i2}(t)$, corresponding to two
different sets of response system parameters, is a measure of the
degree of coincidence between both waveforms. This means that if
two sets of slightly different response system parameters have the
same eye apertures, computed along a limited time period, the
corresponding waveforms are practically equal during this time.

On the contrary, the parameter values of Fig.~\ref{fig:error}~(d)
correspond to the example of \cite{Wang04}, with parameter values
$r^*_4=45.601$ and $\sigma^*_4=15.999$, that undergo a guessing
error on the fifth significative digit. Such error was considered
in \cite{Wang04} unacceptable for correct plaintext recovery.
Effectively, it can be seen in Fig.~\ref{fig:error}~(d) that
$x_r(t)$ and $x(t)$ waveforms are not correlated at all.

If a greater precision in the parameter determination is needed,
the period of measure could be accordingly enlarged. The maximum
allowable precision is limited by the lifespan of the intercepted
communication. To get an infinite precision an infinite measure
period time will be needed. The first steps take very small time
to compute, because the involved number of samples is short, but
the last step is much more time consuming, because the involved
number of samples is very large.

When dealing with very long encrypted messages it may be
unpractical to expand the parameter computation time to the whole
message length, because the computation time may become huge. It
is better to divide the message in fractions of not more than the
equivalent of 1000 s of the Lorenz system, and repeat the
parameter determination procedure for each fraction. In that way
it may happen that the found parameters will be different for each
message fraction.

Once the best values of $r^*$ and $\sigma^*$ are determined, the
plaintext can be retrieved in the same way as the legal key owner
does.

\section{Application to cryptanalysis of a two separated
channels projective synchronization cryptosystem}
\label{twochannel2}

In a recent paper Li and Xu \cite{Li04} proposed a secure
communication scheme based on PS chaotic masking. They illustrated
the feasibility of the scheme with two examples, one of them was
based on the Lorenz system, with sender variables $x_s(t), y_s(t)$
and $z(t)$. The transmitted signals were the Lorenz system shared
scalar variable $z(t)$ and the ciphertext signal, defined as
$U(t)=x_s(t)+ y_s(t)+ m(t)$, where $m(t)$ was the plaintext. The
retrieved plaintext was calculated by the authorized receiver as
$m(t)=U(t)-(x_r(t)+y_r(t))/\alpha$, where $\alpha$ is the PS
scaling factor and $x_r(t)$, $y_r(t)$ are the variables generated
by the response system. The authors claimed that the lack of
knowledge of the value of $\alpha$ by an intruder was an important
feature to assure the information security. In their example the
system parameter values were $\{\sigma, r, b\}=\{10, 60, 8/3\}$,
the scaling factor was $\alpha=5$ and the plaintext was the sound
signal coming from a water flow, of unknown frequency spectrum and
about 0.2 units of amplitude, i.e. approximately 0.005 times the
amplitude of $x_r(t)+y_r(t)$.

We simulated this cryptosystem with arbitrarily chosen sender
initial conditions $x_s(0)=3$, $y_s(0)=3$, $z(0)=20$ because there
was no details about them in~\cite{Li04}. The intruder response
system initial conditions were chosen equal to corresponding
sender initial conditions times the desired scaling factor
$\alpha=5$, that is: $x_r(0)=15$ and $y_r(0)=15$. As plaintext
message was chosen the function $m(t)=0.2 ~\sin(2 \pi~ 30 ~t)$,
i.e. a low frequency tone of similar amplitude to the authors
example.

To break this scheme the same determination procedure described in
the precedent section was employed. First, using the algorithm of
Sec.~\ref{reduction}, it was found that the fixed point $z$
coordinate was $z^*_{C^\pm}=58.9766$, that corresponds to
$r^*=59.9766$ (very close to the true value $r=60$); hence a
practical search range of $r^*$ going from $r^*=59.8$ to
$r^*=60.2$ was selected, equivalent to an error allowance of
$\pm0.33\%$, compliant with Fig.~\ref{fig:r-error} error margins.
The search space of $\sigma^*$, according to Eq. (\ref{eq:sigma}),
should be comprised in the range $0 < \sigma^* <57$.

\begin{figure}[t]
\begin{center}
\psfrag{Eye x aperture}{\scriptsize Eye $x$ aperture}
\begin{overpic}[scale=1]{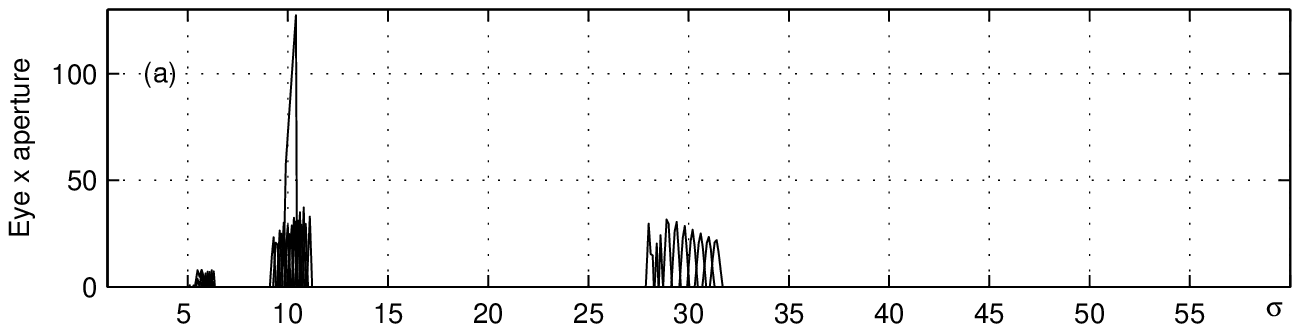}
\end{overpic}
\begin{overpic}[scale=1]{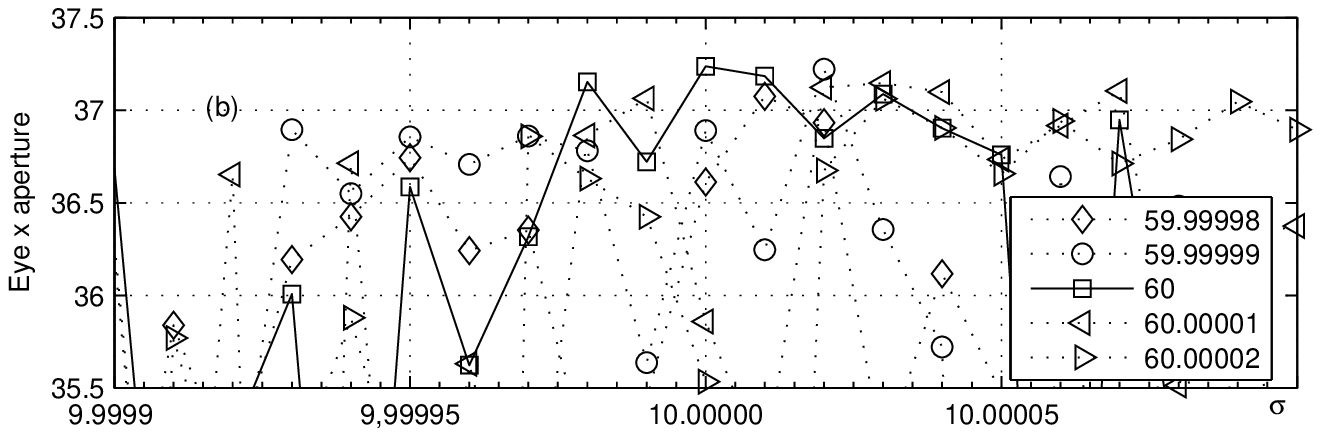}
\end{overpic}
\caption{\label{fig:s1} Intruder receiver eye aperture $x_r$ for
various measure periods: (a) 8 s, with $r^*=59.8$ to $r^*=60.2$;
(b) 800 s, with $r^*=59.9999$ to $r^*=60.0001$.}
\end{center}
\end{figure}

Figure \ref{fig:s1} illustrates the first and fifth steps of the
$r^*$ and $\sigma^*$ determination procedure, that was
accomplished with the same method described in the precedent
section. In the first step, the eye aperture of the receiver $x_r$
variable was measured along a period of 8 s, varying $r^*$ from
$r^*=59.8$ to $r^*=60.2$, and $\sigma^*$ from $\sigma^*=0$ to
$\sigma^*=57$. The result is illustrated in Fig.~\ref{fig:s1}~(a).
As in the previous section it was supposed that the available
ciphertext had a length of 800 s. In Fig.~\ref{fig:s1}~(b) it can
be seen that the discrimination limit of the identification method
was reached for that period of measure, giving multiple peaks
approximately the same eye aperture.

The four peaks of greater amplitude suggest four sets of plausible
potential candidates of response system parameter sets. The
greatest of them, with an eye aperture $x_{a0}= 37.25$, is the
right one: $r^*_0=r=60$, $\sigma^*_0=\sigma=10$; the three
following candidates, in descending order of eye aperture, are
slightly inexact, differing in the seventh significative digit
from the right value: $r^*_1=59.99999$, $\sigma^*_1=10.00002$
($x_{a1}= 37.23$); $r^*_2=60$, $\sigma^*_2=10.00001$ ($x_{a2}=
37.18$); and $r^*_3=60$, $\sigma^*_3=9.99998$ ($x_{a3}= 37.15$).

The determination of the approximated value of the scaling factor
$\alpha^*$ may be achieved by dividing, sample by sample, a time
period $T$ of the ciphertext by the corresponding period of response
system sum of variables and taking the average along that time
period:
\begin{align}
\alpha^* & = \overline{\left( \frac{x_s(t)+y_s(t)+
m(t)}{x_r(t)+y_r(t)}
\right)} \nonumber\\
& = \overline{\left( \frac{x_s(t)+y_s(t)}{x_r(t)+y_r(t)} \right)}+
\overline{\left( \frac{m(t)}{x_r(t)+y_r(t)} \right)} \label{alpha}
\end{align}
where $\overline{f(t)}$ denotes the temporal average of $f(t)$
over a period T. In the case that $m(t)$ has zero mean, as in the
example given in \cite{Li04}, the last quotient vanishes since
$m(t)$ is independent of $x_r(t)+y_r(t)$, while the mean of the
first quotient of Eq. \ref{alpha} reveals the value of $\alpha^*$.
This simple procedure may be slightly inexact due to
divide-by-cero problems, so the low amplitude samples were
eliminated and the following algorithm was used to determine
$\alpha^*$ with more accuracy:

\begin{enumerate}
    \item select a collection of samples of $x_r(t)$ and $y_r(t)$,
    corresponding to the 800 first seconds of the waveform,
    \item calculate the maximum value $M_{x+y}$ of the
    collection of $|x_r(t)+y_r(t)|$ samples,
    \item compile a list of all the exact sampling times $t_{js}$
    for which $|x_r(t_{js})+y_r(t_{js})|>0.3~M_{x+y}$ and count the number of
    them $ns$,
    \item calculate the scaling factor as $\alpha^*=\frac{1}{{ns}}
    \sum\limits_1^{ns} {\frac{{x_r(t_{js})+y_r(t_{js})}}{{U(t_{js} )}}}$.
\end{enumerate}

\begin{figure}[t]
\psfrag{Retrieved text amplitude}{\scriptsize Plaintext amplitude}
\begin{overpic}[scale=1]{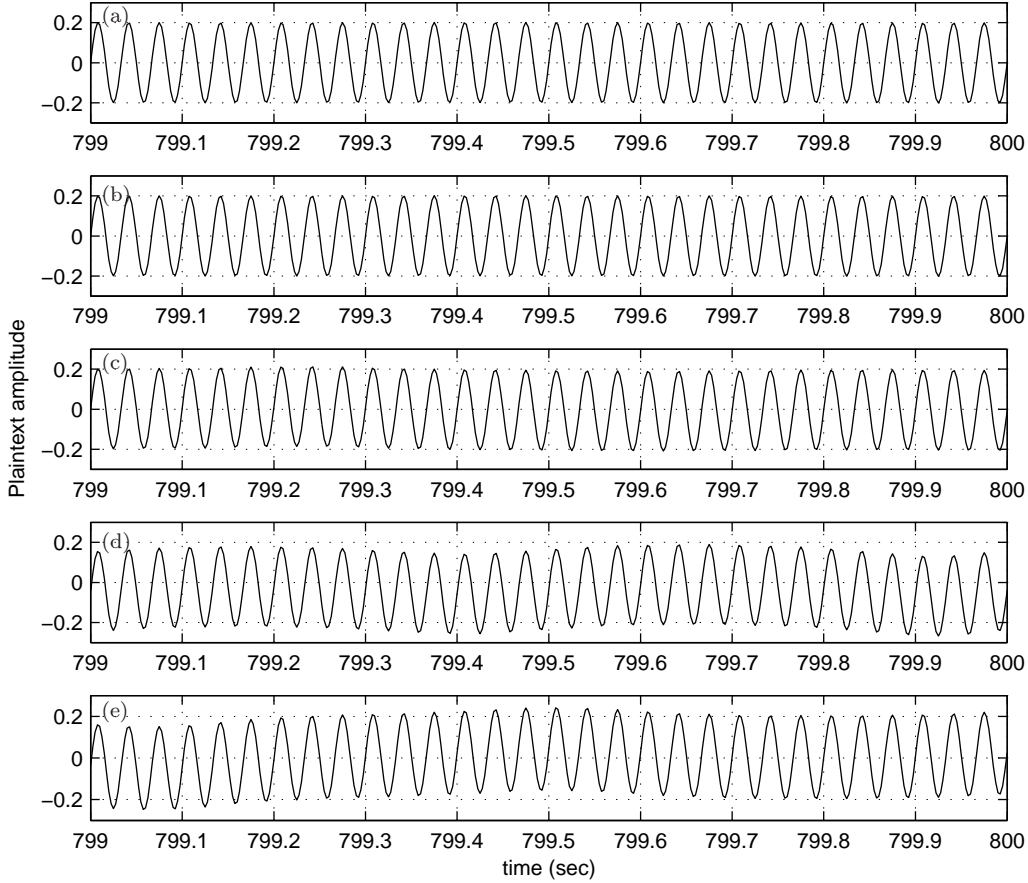}
     \put(10.5, 83) {\scriptsize (a)}
     \put(10.5, 66) {\scriptsize (b)}
     \put(10.5, 50) {\scriptsize (c)}
     \put(10.5, 33) {\scriptsize (d)}
     \put(10.5, 17) {\scriptsize (e)}
\end{overpic}
\caption{\label{fig:diference} Last second of plaintext. (a)
Original message. Retrieved plaintext for four sets of response
system parameters: (b) $r^*_1=60$, $\sigma^*_1=10$; (c)
$r^*_1=59.99999$, $\sigma^*_1=10.00002$; (d) $r^*_2=60$,
$\sigma^*_2=10.00001$; (e) $r^*_3=60$, $\sigma^*_3=9.99998$.}
\end{figure}

The result was $\alpha ^* = 5.000038$, for all the four parameter
sets previously identified, which represent a relative error of
$7\times10^{-6}$ related to $\alpha$, that will affect the
recovering of $m(t)$ adding a negligible noise of 63 db below its
amplitude.

The retrieved plaintext was then calculated as:
\begin{equation}
m^*(t)= U(t)-\frac{x_r(t)+y_r(t)}{ \alpha^*}
=x_s(t)+y_s(t)+m(t)-\frac{x_r(t)+y_r(t)}{ \alpha^*}
\end{equation}

The Fig.~\ref{fig:diference} illustrate the plaintext waveform
between the seconds 799 and 800 of the the original message $m(t)$
and four recovered messages $m^*(t)$, for the four system
parameter sets previously identified.  It can be seen that the
retrieved waveform corresponding to the first and second set of
intruder receiver parameters is completely equal to the original
plaintext, while the third and fourth parameters sets cause a
small distortion of the retrieved plaintext; note that the
distortion increases as the eye aperture goes down, as expected.
Nevertheless any of the four potential candidates of response
system parameter sets may be used indistinctly to gain access to
the encrypted information without significant error, for the
limited time period that was considered for their determination.

\section{Simulations}

All results were based on simulations with MATLAB 7, the Lorenz
integration algorithm was a four-fifth order Runge-Kutta with an
absolute error tolerance of $10^{-9}$, a relative error tolerance of
$10^{-6}$, and a sampling frequency of 400 Hz.

\section{Conclusion}
\label{Sec:Conclusion}

This work describes a novel Lorenz system parameter determination
procedure, based on the measure of some attractor geometrical
properties, with the help of a homogeneous driving synchronization
mechanism. The method is applicable to the cryptanalysis of a
two-channel chaotic cryptosystem that uses the variable $z$ as a
synchronization signal, allowing for the system secret key
recovery and evincing that such systems are not suitable for
secure communications. The method is not applicable to break
two-channel chaotic cryptosystems that use the variable $x$ or $y$
as a synchronization signal.

\section*{Acknowledgements}This work was supported by Ministerio de Ciencia y
Tecnolog\'{\i}a of Spain, research grant SEG2004-02418, and by The
Hong Kong Polytechnic University's Postdoctoral Fellowships Scheme
under grant no. G-YX63.

\end{document}